\documentclass[aip,jcp,preprint]{revtex4-1}

\usepackage{graphics}
\usepackage{graphicx}
\usepackage{amsfonts}
\usepackage{textcomp}
\usepackage{amssymb}
\usepackage{amsmath}
\usepackage{rotating}
\usepackage{multirow}
\usepackage{makecell}
\usepackage{hyperref}

\begin{document}

\title{Inelastic Neutron Scattering Analysis with Time-Dependent Gaussian-Field Models}

\author{Cedric J. Gommes} 
\altaffiliation{Current affiliation: Department of Chemical Engineering, University of Li\`ege B6A, all\'ee du Six Ao\^ut 3, B-4000, Li\`ege, Belgium} \email{cedric.gommes@uliege.be}
\affiliation{Forschungszentrum J\"ulich GmbH, J\"ulich Center for Neutron Science, 52425 J\"ulich, Germany}
\author{Reiner Zorn}
\affiliation{Forschungszentrum J\"ulich GmbH, J\"ulich Center for Neutron Science, 52425 J\"ulich, Germany}

\author{Sebastian Jaksch}
\author{Henrich Frielinghaus}
\author{Olaf Holderer}
\affiliation{Forschungszentrum J\"ulich GmbH, J\"ulich Center for Neutron Science at the Heinz Maier Leibnitz Zentrum, Lichtenbergstrasse 1, 85747 Garching, Germany}

\date{\today}

\begin{abstract}
Converting neutron scattering data to real-space time-dependent structures can only be achieved through suitable models, which is particularly challenging for geometrically disordered structures. We address this problem by introducing time-dependent clipped Gaussian field models. General expressions are derived for all space- and time-correlation functions relevant to coherent inelastic neutron scattering, for multiphase systems and arbitrary scattering contrasts. Various dynamic models are introduced that enable one to add time-dependence to any given spatial statistics, as captured {\it e.g.} by small-angle scattering.  In a first approach, the Gaussian field is decomposed into localised waves that are allowed to fluctuate in time or to move, either ballistically or diffusively. In a second approach, a dispersion relation is used to make the spectral components of the field time-dependent. The various models lead to qualitatively different dynamics, which can be discriminated by neutron scattering. The methods of the paper are illustrated with oil/water microemulsion studied by small-angle scattering and neutron spin-echo. All available data - in both film and bulk contrasts, over the entire range of $q$ and $\tau$- are analyzed jointly with a single model. The analysis points to static large-scale structure of the oil and water domains, while the interfaces are subject to thermal fluctuations. The fluctuations have an amplitude around 60 \AA \ and contribute to 30 \% of the total interface area.
\end{abstract}


\maketitle


\section{Introduction}

Neutron scattering is one of the few experimental techniques that allow one to probe both the structure and the dynamics of physical systems at the {\r A}ngstr{\"o}m scale\cite{Sivia:2011,Squires:2012}. Typically, structural information is obtained through the elastic scattering of cold or thermal neutrons (SANS). The dynamic information is obtained through inelastic and quasi-elastic scattering, as the neutrons gain or lose energy when they interact with moving phases in the system. As for most scattering techniques, however, converting experimental data to real-space and time-dependent structures can be challenging. This is particularly the case for complex and disordered structures that cannot be described in simple geometrical terms. 

When studying disordered systems, stochastic models often provide a practical compromise between geometrical realism and mathematical simplicity. The former is necessary to account for as many geometrical features as possible, and the latter improves the robustness of the analysis by avoiding unnecessarily large number of parameters \cite{Serra:1982,Torquato:2002,Lantuejoul:2002}. In that spirit, stochastic models have often been used to analyze small-angle scattering data from a variety of physical systems and reconstruct their structure \cite{Sonntag:1981,Roberts:1997,Gille:2011,Gommes:2018,Prehal:2020}. In the present paper, we generalize this type of approach to analyze and model time-dependent structures investigated by inelastic neutron scattering.

The paper focuses specifically on a family of descriptive models based on clipped Gaussian random fields. These models originate in the work of Cahn on spinodal decomposition \cite{Cahn:1965}, but they have since been used as general geometrical models of disordered structures in a variety of contexts, including porous materials \cite{Quiblier:1984,Roberts:1995,Gommes:2018}, polymers \cite{Chen:1996,DHollander:2010}, emulsions \cite{Berk:1987,Teubner:1991}, gels \cite{Gommes:2008}, confined liquids \cite{Gommes:2013,Gommes:2018B}, nanoparticles \cite{Gommes:2020}, etc. Gaussian random fields are comprehensively characterized by their correlation function, which makes them particularly useful in the context of scattering studies. 

The theoretical developments of the present paper are illustrated on previously-published elastic and inelastic neutron scattering data measured on water/oil microemulsion, which are presented shortly in Section II, together with some general results pertaining to elastic and inelastic neutron scattering. Section III covers some classical results of static Gaussian-field models, which are generalized to time-dependent structures in Section IV. Three families of dynamic models are proposed, which are applicable to any static Gaussian field and endow it with qualitatively different time-dependence. In Section V, some aspects of the models relevant to inelastic scattering are discussed, and the models are used to analyze the microemulsion data.

\section{Neutron small-angle scattering and spin-echo data}
\label{sec:experimental}

\begin{figure}
\begin{center}
\includegraphics[width=10cm]{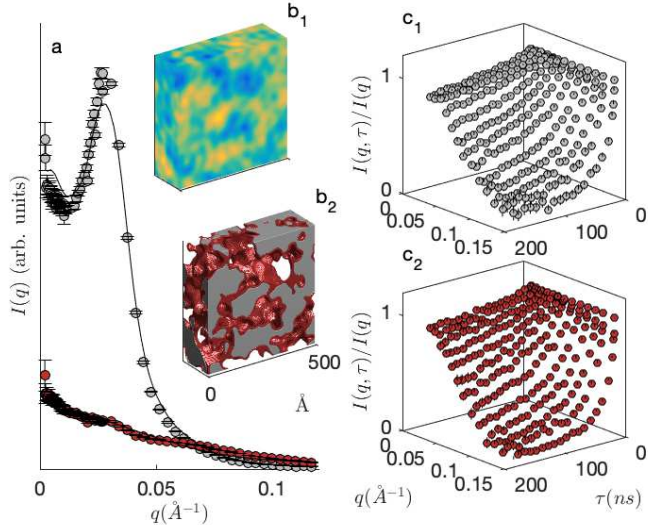}
\caption{Neutron Small-Angle Scattering data (SANS, $a$) measured on a microemulsion in bulk (grey) and film (red) contrasts, together with structure reconstructed from it as a clipped Gaussian field model ($b_1$: field, $b_2$ clipped structure with oil in grey and surfactant in red). The neutron spin-echo (NSE) data measured in the same conditions are shown in $c_1$ and $c_2$. In the SANS patterns (a) the dots are the experimental values, and the solid lines are the fitted model. The values in the Gaussian field shown in $b_1$ range from -2.5 (blue) to + 2.5 (yellow). The error bars are $\pm 2 \sigma$ for both SANS and NSE.}
\label{fig:data}
\end{center}
\end{figure}

The methods and models developed in the paper are illustrated with published neutron small-angle scattering (SANS) patterns and neutron spin-echo (NSE) data measured on a water/oil microemulsion stabilised with a surfactant\cite{Holderer:2005,Holderer:2007}. The relevant data are available on the authors institutional repository\cite{Holderer:2021} and they are displayed in Fig. \ref{fig:data}. 

The bicontinuous phases of the microemulsion consisted in water and decane with decyl-polyglycol-ether (C$_{10}$E$_{4}$) as a surfactant. The volume fractions of the three phases were $\phi_o \simeq 0.4075$, $\phi_s \simeq 0.185$ and $\phi_a \simeq 0.4075$ for oil, surfactant and aqueous phases, respectively. Small amounts (0.25 wt.\%) of homopolymers were dispersed in the continuous phases in order to slightly modify their viscosity and the efficiency of the surfactant (see Refs. \cite{Holderer:2005,Holderer:2007}), namely polyethylene oxide (PEO) in water and polyethylene propylene (PEP) in decane. The molecular weights slightly differed in SANS (10 kg/mol) and NSE (5 kg/mol) experiments, which has only minor effects for the purposes of this study on the relaxation rate in the NSE experiments ($<10$ \%), but gave a complete set of SANS and NSE data. The microemulsion was prepared in two different neutron scattering contrasts, by exchanging hydrogen with deuterium. In the so-called bulk contrast, deuterated water (D$_2$O) was used with protonated surfactant and oil, which results in a contrast between the water domains and the oil-surfactant-domains. In film contrast, the decane was deuterated as well, leaving only the protonated surfactant film visible in the deuterated water/oil surrounding. The SANS experiments were conducted on the KWS-2 small angle scattering instrument at the DIDO reactor of Forschungszentrum J\"ulich, the NSE experiments were conducted on the IN15 instrument at the Institut Laue-Langevin in Grenoble. The resolutions the SANS and NSE data in Fig. \ref{fig:data} are $\sigma^{SANS}_q=0.0034$ \AA$^{-1}$ and $\sigma^{NSE}_q=0.0085$ \AA$^{-1}$, respectively.

Microemulsions are strong coherent scatterers, so that incoherent scattering from individual atoms (mainly hydrogen) does not play a significant role at the length scales discussed in this paper. Therefore, the central structural characteristic of the microemulsion relevant to both the SANS and NSE data is the scattering-length correlation function\cite{VanHove:1954,Squires:2012}
\begin{equation} \label{eq:C_rho_def}
C_{\rho}(r,\tau) = \langle \rho( \mathbf{x},t)  \rho(\mathbf{x}+ \mathbf{r},t+\tau) \rangle - \langle \rho \rangle^2
\end{equation}
which characterises the statistical correlation between the scattering length density $\rho$ at two points at a distance $r$ apart, and time lag $\tau$. Throughout the paper we assume statistical isotropy, so that correlation functions depend only on the modulus of the distance $r=|\mathbf{r}|$. In Eq. (\ref{eq:C_rho_def}) the brackets $\langle \rangle$ stand for the average value, evaluated over all accessible positions $\mathbf{x}$ and times $t$. For the type of ergodic models considered later in the paper, they can also be thought of as ensemble averages.\cite{Torquato:2002,Lantuejoul:2002,Gommes:2018} 

When working with stochastic models it is convenient to introduce the concept of covariance \cite{Serra:1982,Lantuejoul:2002}, which is occasionally also referred to as 2-point probability functions\cite{Torquato:2002} or stick-probability functions\cite{Ciccariello:1981}. The covariance of, say the oil phase $o$ of the microemulsion, is defined as the probability for two points at distance $\mathbf{r}$ from one another to belong to that phase at two moments separated with time lag $\tau$, namely
\begin{equation}
C_{oo}(r,\tau ) =\textrm{Prob} \left[ \left(\mathbf{x} \in o \ \textrm{at time} \ t \right) \& \left( \mathbf{x} + \mathbf{r} \in o \ \textrm{at time} \ t+\tau \right) \right]
\end{equation}
As this generalizes to {\it cross-covariances} for two points belonging to two distinct phases, the name {\em self-covariance} is occasionally used to insist that the two points belong to the same phase.  Because each of the three phases of the microemulsion  - oily, aqueous and surfactant - has a specific scattering-length density, the correlation function $C_\rho(r)$ is a linear combination of the covariances of the phases. Out of the six self- and cross-covariances that are defined for a three-phase system, only three are linearly independent.\cite{Torquato:2002} A convenient expression for $C_\rho $ is therefore \cite{Gommes:2013}
\begin{align} \label{eq:C_rho}
C_{\rho}(r,\tau) &= (\rho_o - \rho_s)(\rho_o-\rho_a) [C_{oo}(r,\tau) - \phi_o^2]+ (\rho_s - \rho_o)(\rho_s-\rho_a) [C_{ss}(r,\tau) - \phi_s^2]\cr
&+(\rho_a - \rho_o)(\rho_a-\rho_s) [C_{aa}(r,\tau) - \phi_a^2]
\end{align}
where $\rho_o$, $\rho_s$ and $\rho_a$ are the scattering-length densities of the oil, surfactant, and aqueous phases, respectively; $C_{oo}$, $C_{ss}$ and $C_{aa}$ are the corresponding self-covariances. A derivation of Eq. (\ref{eq:C_rho}) is provided in the Supplementary Material (Sec. SM-1). The bulk contrast relevant to Fig. \ref{fig:data} correspond to $\rho_o=\rho_s \neq \rho_a$, in which case $C_\rho$ is proportional to $C_{aa}$. The film contrast corresponds to $\rho_o=\rho_a \neq \rho_s$, and in that case $C_\rho$ is proportional to $C_{ss}$. 

The coherent inelastic neutron scattering data is expressed in terms of the intermediate scattering function $I(q,\tau)$. The latter is defined as the Fourier transform of the correlation function $C_\rho(r,\tau)$, namely \cite{Sivia:2011,Squires:2012}
\begin{equation} \label{eq:I_Fourier}
I(q,\tau) = \int_0^\infty \frac{\sin(qr)}{qr} C_{\rho}(r,\tau) 4 \pi r^2 \textrm{d}r
\end{equation}
and the instrument resolution is accounted by multiplying $C_\rho$ by a spread function with width $\sigma_q$, prior to Fourier transform. The situation relevant to SANS is elastic scattering corresponding to $I(q,0)$, to which we refer simply as $I(q)$ when there is no ambiguity. The data measured in NSE instruments is $I(q,\tau)/I(q)$, as given in Fig. \ref{fig:data}c1 and \ref{fig:data}c2 for the microemulsion.

In Fig. \ref{fig:data} the SANS data in both film and bulk contrasts were fitted jointly with a clipped Gaussian field model, adapting a procedure developed elsewhere\cite{Gommes:2018}. For the sake of completeness, the detailed procedure is described in the Supplementary Material (Sec. SM-3.3).  A realisation of the model is shown in Fig. \ref{fig:data}b.

\section{Clipped Gaussian-field models}
\label{sec:static}

\subsection{Static Gaussian random fields}
\label{sec:GRF}

We focus here on static, that is time-independent, Gaussian random fields (GRF) and we introduce time-dependence in Sec. \ref{sec:timedependentmodels}. A convenient and classical way to think of GRFs is as a superposition of random sine waves \cite{Berk:1991,Levitz:1998}
\begin{equation} \label{eq:W_def}
W(\mathbf{x}) = \sqrt{\frac{2}{N}} \sum_{n=1}^N \sin\left[ \mathbf{q}_n \cdot \mathbf{x} - \varphi_n \right]
\end{equation}
where the phases are uniformly distributed over $[0, 2 \pi)$ and the wavevectors $\mathbf{q}$ are drawn from a user-specified density distribution over reciprocal space $f_W(\mathbf{q}) \textrm{d}V_q$, referred to as the spectral density of the field. For asymptotically large values of $N$, the central limit theorem ensures that $W(\mathbf{x})$ is Gaussian-distributed at any point $\mathbf{x}$ with average equal to zero, and the factor in Eq. (\ref{eq:W_def}) ensures that the variance is equal to one. 

A central characteristic of the GRF in the context of elastic scattering is its correlation function $g_W(r)$, defined as the statistical correlation between the values of $W(\mathbf{x})$ at two points at distance $r$ apart
\begin{equation} \label{eq:gW}
g_W(r) = \langle W(\mathbf{x}) W(\mathbf{x}+\mathbf{r}) \rangle
\end{equation}
where the brackets have the same meaning as in Eq. (\ref{eq:C_rho_def}), and the dependence is only on the modulus $r =|\mathbf{r}|$ for isotropic fields. The field correlation function is obtained as the Fourier transform of the spectral density, namely\cite{Berk:1987,Berk:1991}
\begin{equation} \label{eq:gW_f}
g_W(r) = \int_0^\infty \frac{\sin(qr)}{qr} f_W(q) 4 \pi q^2 \textrm{d}q
\end{equation}
In principle any integrable and positive function can be used as a spectral density. In practice, in order to ensure that the structures modelled by clipping the field have finite surface areas\cite{Berk:1991,Teubner:1991}, it is necessary to impose that the second moment of $f_W(q)$ be finite. This enables one to define $l_W$ as
\begin{equation} \label{eq:lW}
\frac{1}{l_W^2} = \frac{1}{6} \int_0^\infty q^2 f_W(q) 4 \pi q^2 \textrm{d}q
\end{equation}
which we refer to as the field characteristic length. The finiteness of $l_W$ corresponds to a quadratic behavior of the correlation function $g_W \simeq 1 - (r/l_W)^2 + \ldots $ for small distances, and is a condition for the modelled structures to have finite surface areas\cite{Teubner:1991,Berk:1991}. 

\begin{sidewaystable}
\caption{Examples of static Gaussian random fields (GRFs) with their spectral densities $f_W(q)$, field correlation functions $g_W(r)$, and characteristic lengths $l_W$. The function $w(|\mathbf{x}|)$ is the corresponding elementary wave relevant to a dilution approach (see text, $l$ and $\mu$ are model parameters). These functions are plotted in Figs. SM-1 to SM-8 of the Supplementary Material.}
\begin{center}
\begin{tabular}{c|c|c|c|c|r}
GRF\#  & $f_W(q)$ & $g_W(r)$ &  $l_W$  & $w(\mathbf{x})$$^a$ & Ref. \cr
\hline
1 
&  $ \frac{l^2}{16 \pi^3} \delta[q - \frac{2 \pi}{l}]  $ 
& $\frac{\sin[2\pi r/l]}{2\pi r /l} $ & $\sqrt{6} l/(2 \pi) $ 
& -$^{b}$  
& \cite{Berk:1987} \cr
2
&   $\left( \frac{l}{2 \sqrt{\pi}} \right)^3   e^{-[q l]^2/4}  $ 
& $e^{-[r/l]^2}$ 
& $l$ 
& $ e^{- 2 \left( \frac{|\mathbf{x}|}{l} \right)^2} $   
& \cite{Lantuejoul:2002,Gelfand:2010} \cr
3 
&   $ \frac{l^2}{4q} \frac{\sinh[\pi ql/2]}{1 + \cosh[\pi q l]}  $ 
& $\frac{1}{\cosh[r/l]} $ & $\sqrt{2} l $ 
& -$^{b}$  
& \cite{Gommes:2008} \cr
4  
&  $\left(\frac{l}{\sqrt{\pi}} \right)^3 \frac{1}{480} (ql)^4 e^{-[ql]^2/4} $
& $\left[1 - \frac{4}{3}  \left( \frac{r}{l} \right)^2 + \frac{4}{15} \left( \frac{r}{l} \right)^4 \right] e^{ -\left( \frac{r}{l} \right)^2 } $ 
&   $\sqrt{3/7} l $ 
& $ \left[  \left(\frac{ |\mathbf{x}|}{l}\right)^2 - \frac{3}{4} \right] e^{- 2 \left(\frac{|\mathbf{x}|}{l} \right)^2} $  
& \cr
5  
&  $ \frac{l^3}{4\pi^2 \mu} \frac{ \sinh[\pi^2/\mu] \sinh[\pi ql/(2\mu)]/(ql)}{\cosh[2 \pi^2 /\mu] + \cosh[\pi ql/ \mu]}  $ 
& $\frac{\sin[2\pi r/l]}{(2\pi r /l)\cosh[\mu r/l]} $ 
& $ l /\sqrt{ \frac{2 \pi^2}{3} + \frac{\mu^2}{2} } $ 
& -$^{b}$ 
&  \cite{Gommes:2008} \cr
6
&  $ \left(\frac{l}{\sqrt{\pi}}\right)^3 \frac{\Gamma(\mu+\frac{3}{2})}{\Gamma(\mu) [1 + (ql)^2]^{\frac{3}{2}+\mu}} $ 
& $2^{1-\mu}  \frac{(r/l)^\mu K_\mu(r/l)}{ \Gamma(\mu + 1)}$ 
&  $ 2l \sqrt{\mu - 1} $ 
&  $\left(\frac{|\mathbf{x}|}{l} \right)^{\frac{\mu}{2}-\frac{3}{4}}  K_{\frac{\mu}{2}-\frac{3}{4}} \left(\frac{|\mathbf{x}|}{l} \right)$
& \cite{Gelfand:2010,Lantuejoul:2002} \cr
7  
&  $ \left(\frac{l}{\sqrt{\pi}}\right)^3 \frac{\Gamma(\mu+1) }{\Gamma(\mu-\frac{1}{2})} [1 - (ql)^2]^{\mu-\frac{3}{2}} $ 
& $2^\mu \Gamma(\mu + 1) \frac{J_\mu(r/l)}{(r/l)^\mu}$ 
&  $ 2l \sqrt{\mu+1}$ 
&  $\left(\frac{|\mathbf{x}|}{l} \right)^{-\frac{\mu}{2}-\frac{3}{4}}  J_{\frac{\mu}{2}+\frac{3}{4}} \left(\frac{|\mathbf{x}|}{l} \right)$
& \cite{Lantuejoul:2002} -$^c$ \cr
8
& Piecewise Linear 
& Cf. Sup. Info.
& Cf. Sup. Info.
& -$^b$   
& \cite{Gommes:2018} \cr
\end{tabular}
\end{center}
\label{tab:fields}
$^a$: within unspecified normalizing factor; $^b$: not available; $^c$: a typo in the formula provided for $f_W(q)$ has been corrected in the present table. 
\end{sidewaystable}%

The methods developed in the paper apply to any static Gaussian field. A few examples are given in Tab. \ref{tab:fields} with explicit spectral densities, correlation functions, and characteristic lengths $l_W$, which are also plotted in Figs. SM-1 to SM-8 of the Supplementary Material. These fields are referred to in the rest of the paper by the number in the first column. GRF-1 contains a single spectral component, and is arguably the simplest possible Gaussian field. By contrast, GRF-2 is extremely polydispersed and is referred to in geostatistics as the squared-exponential correlation function. GRF-3 is also polydispersed: its correlation function is exponential for asymptotically large distances, but the $1/\cosh$ function ensures quadratic shape at the origin and hence finite $l_W$. GRF-4 is introduced in Sec. \ref{sec:data_analysis}, and leads to structures with a scattering peak. GRF-5 is obtained by multiplying the correlation functions of GRF-1 and GRF-3, and provides one with a parameter to control the polydispersity of the structure, which makes it convenient for SAS data fitting\cite{Gommes:2008,Prehal:2017}. Other examples discussed in the scattering literature can be found {\it e.g.} in Refs. \cite{Teubner:1987,Chen:1996,Roberts:1997}. The following two entries in Tab. \ref{tab:fields} are classical in geostatistics but are seldom used in scattering studies. GRF-6 is the Mat\'ern model where $K_\mu$ is a modified Bessel function. The parameter $\mu$ controls the smoothness of the field, which is $\mu-1$ times differentiable\cite{Rasmussen:2006}, and GRF-2 is obtained as a particular case for $\mu \to \infty$. By contrast to GRF-6, field GRF-7 introduces strong correlations through Bessel function $J_\mu$. It leads to peaked scattering functions (see Fig. SM-7) and coincides with GRF-1 in the limit $\mu \to 1/2$. 

When it comes to analyzing experimental scattering patterns, the simple analytical expressions in Tab. \ref{tab:fields} seldom provide sufficient flexibility for data fitting. Therefore, a convenient approach consists in linearly combining independent Gaussian fields $W_i(\mathbf{x})$, with spectral densities $f_W^{(i)}(q)$, so as to create a composite field
\begin{equation} \label{eq:combine_W}
W(\mathbf{x}) =\sum_i \sigma_i W_i(\mathbf{x})
\end{equation}
where $\sigma_i$ are constants. The spectral density of the resulting field is
\begin{equation}
f_W(q) =\sum_i \sigma^2_i f_W^{(i)}(q)
\end{equation}
and a similar relation holds for $g_W(r)$. Because the integral of $f_W(q)$ over the entire reciprocal space is the variance of the field, the parameter $\sigma_i^2$ can be thought of as the contribution of $W_i(\mathbf{x})$ to the total variance of the composite field $W(\mathbf{x})$. In that spirit, a possible approach to data fitting would consist in combining a large number of monodispersed fields ({\it e.g.} GRF-1 in Tab. \ref{tab:fields}), so as to approximate an experimental spectral density as a sum of Dirac peaks. As an unpractically large number of peaks might be needed to approximate a continuous function, a more practical approach consists in replacing the Dirac peaks by broader functions. The piecewise-linear model (GRF-8 in Tab. \ref{tab:fields}) corresponds to such an approach, which was developed in earlier work\cite{Gommes:2018}. As this approach was used here to fit the SANS data in Fig. \ref{fig:data}a, it is described in detail in the Supplementary Material (Sec. SM-3). In particular, the influence of the number of nodes for the SANS fit shown in Fig. \ref{fig:data}a is illustrated in Fig. SM-10.

When generalising the Gaussian-field modelling to time-dependent structures it will prove useful to use another construction of Gaussian fields, which is mathematically equivalent to Eq. (\ref{eq:W_def}). In so-called dilution random functions, \cite{Serra:1982,Lantuejoul:1991,Lantuejoul:2002} a field is created as a sum of localised elementary waves $w(\mathbf{x})$, randomly positioned in space, namely
\begin{equation} \label{eq:W_def_dilution}
W(\mathbf{x}) = \sum_s A_s w(\mathbf{x} - \mathbf{x}_s)
\end{equation}
where the sum is on all the seeds $\mathbf{x}_s$ of a Poisson point process with density $\theta$, and $A_s$ is any random amplitude satisfying $\langle A \rangle = 0$ and $\langle A_s A_{s'} \rangle = \langle A^2 \rangle \delta_{ss'}$. The latter condition corresponds to uncorrelated wave amplitudes. In the limit of a large density of the Poisson process, many elementary waves overlap at any given point of space so that the values of the field defined in Eq. (\ref{eq:W_def_dilution}) become Gaussian distributed. 

In the context of a dilution approach, the correlation function of the field is calculated as\cite{Serra:1982,Lantuejoul:1991,Lantuejoul:2002}
\begin{equation} \label{eq:gW_dilution}
g_W(\mathbf{r})= \theta \langle A^2 \rangle K(\mathbf{r})
\end{equation}
where 
\begin{equation}
K(\mathbf{r})= \int \textrm{d}V_x \  w(\mathbf{x}) w(\mathbf{x} - \mathbf{r}) 
\end{equation}
is the self-convolution of the elementary wave. In order to ensure that the variance of the field is equal to one, one has to impose $g_W(0)=1$, which requires adjusting the amplitudes so that $\theta \langle A^2 \rangle K(0)=1$. Although Eqs. (\ref{eq:W_def}) and (\ref{eq:W_def_dilution}) are conceptually different constructions, the two approaches are mathematically equivalent. The spectral density of the dilution model is indeed obtained as
\begin{equation} \label{eq:fW_w}
    f_W(\mathbf{q}) = \theta \langle A^2 \rangle \left| \int \textrm{d}V_x \  w(\mathbf{x}) \exp ( i \mathbf{q} \cdot \mathbf{x} ) \right|^2
\end{equation}
which results from evaluating the Fourier transform of Eq. (\ref{eq:gW_dilution}). Among the static Gaussian fields presented in Tab. \ref{tab:fields}, the shape of the elementary wave is know for GRF-2, GRF-4, GRF-6 and GRF-7. Conceptually, however, any field such that $\sqrt{f_W(q)}$ is integrable can be thought of as resulting from a superposition of a large number of randomly positioned elementary waves.

\subsection{Clipping procedure}

\begin{figure}
\begin{center}
\includegraphics[width=8cm]{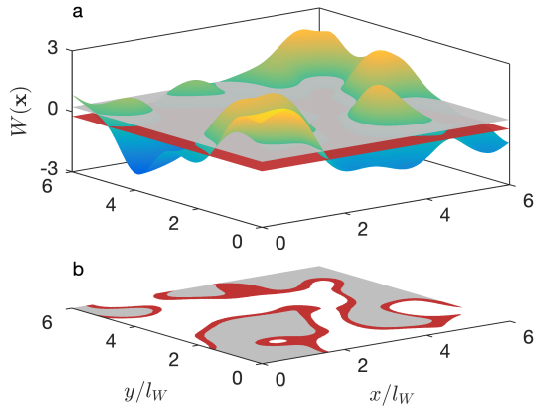}
\caption{Microemulsion modelling as a clipped Gaussian field, with the underlying field shown in (a) and the structure in (b). The clipping thresholds are $\alpha = -0.234$ and $\beta = +0.234$, resulting in aqueous (white) and oil (grey) phases with volume fractions $\phi_a=\phi_o=0.4075$, and $\phi_s=0.185$ for the surfactant (red). The figure is a 2D cut out of a 3D realization obtained from GRF-2 of Tab. \ref{tab:fields}, with distances normalized to $l_W$.}
\label{fig:GRF2D}
\end{center}
\end{figure}

According to a classical approach, the phases of disordered systems can be modelled as excursion sets of a Gaussian field $W(\mathbf{x})$, which is also referred to as clipped-Gaussian-field models.\cite{Quiblier:1984,Berk:1987} In the particular case of emulsions\cite{Teubner:1991} a convenient clipping procedure is based on two thresholds $\alpha \leq \beta$, as sketched in Fig. \ref{fig:GRF2D}. The oil phase is modelled as the points of space where $\beta \le W(\mathbf{x})$, the surfactant film-like phase as $\alpha \le W(\mathbf{x}) < \beta$, and the aqueous phase as $W(\mathbf{x})< \alpha$. 

Because the values of $W(\mathbf{x})$ are Gaussian distributed, the values of the thresholds control the volume fractions of the phases. The volume fraction of the oil $\phi_o$, is obtained as
\begin{equation} \label{eq:phi_o}
\phi_o = \Lambda_1[\beta] 
\end{equation}
where the function $\Lambda_1[x]$ is the probability for a univariate Gaussian variable to take values larger than $x$, which can be calculated as
\begin{equation}
\Lambda_1[x] = \frac{1}{2} \left( 1 - \textrm{erf}[x/\sqrt{2}] \right)
\end{equation}
where erf is the error function. With the same notation, the volume fraction of the surfactant phase is 
\begin{equation} \label{eq:phi_s}
\phi_s = \Lambda_1[\alpha]  - \Lambda_1[\beta] 
\end{equation}
and the volume fraction of the remaining aqueous phase is $\phi_a = 1 - \phi_o - \phi_s$. Relevant values for the microemulsion of Sec. \ref{sec:experimental} are $\alpha=-0.234$ and $\beta \simeq +0.234$, corresponding to $\phi_f \simeq 0.185$ and $\phi_a = \phi_o= 0.4075$, as used in Fig. \ref{fig:GRF2D}.

The scattering functions are obtained from the covariances of the various phases of the microemulsion, which are calculated from the field correlation function $g_W(r)$ and the clipping thresholds $\alpha$ and $\beta$. In line with Eq. (\ref{eq:C_rho}), we consider here the covariances $C_{oo}(r)$, $C_{ss}(r)$ and $C_{aa}(r)$, defined as the probabilities for two randomly chosen points are distance $r$ from one another to belong both to the oil, surfactant, and aqueous phases, respectively. The covariances are expressed in terms of the bivariate error function $\Lambda_2[\alpha,\beta,g]$, defined as the probability for two correlated Gaussian variables, with correlation $g$, to take values larger than $\alpha$ and $\beta$, respectively \cite{Roberts:1995}. Explicitly the expressions are\cite{Teubner:1991,Levitz:1998}
\begin{equation} \label{eq:Coo}
C_{oo}(r) = \Lambda_2[\beta, \beta, g_W(r)]
\end{equation}
for the oil phase, 
\begin{equation} \label{eq:Css}
C_{ss}(r) = \Lambda_2[\alpha, \alpha, g_W(r)]  + \Lambda_2[\beta, \beta,g_W(r)] - 2 \Lambda_2[\alpha, \beta, g_W(r)]
\end{equation}
for the surfactant phase, and
\begin{equation} \label{eq:Caa}
C_{aa}(r) = 1 - 2 \Lambda_1[\alpha]+ \Lambda_2[\alpha, \alpha, g_W(r)] 
\end{equation}
for the aqueous phase. The values used to calculate the scattering functions through Eqs. (\ref{eq:C_rho}) are the centred covariance $\bar C_{oo}(r)$, $\bar C_{ss}(r)$ and $\bar C_{aa}(r)$, obtained by subtracting the corresponding squared volume fraction, so as to enable their Fourier transformation through  Eq. (\ref{eq:I_Fourier}).

\begin{figure}
\begin{center}
\includegraphics[width=5cm]{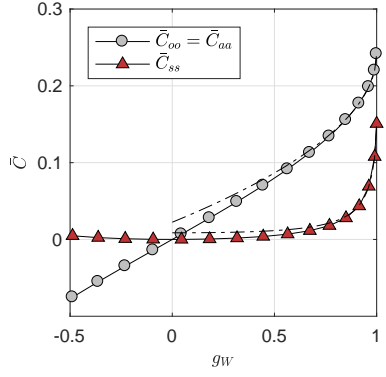}
\caption{Effect of clipping: non-linear relation between the field correlation $g_W$ and the centered covariance $\bar C = C - \phi^2$ of the aqueous, oil, and surfactant phases, calculated from Eqs.  (\ref{eq:Coo}-\ref{eq:Caa}) with clipping constants $\alpha \simeq -0.234$ and $\beta \simeq +0.234$. The relations are highly non-linear for $g_W \simeq 1$ and linear for asymptotically small $g_W$, which limit is relevant for asymptotically large values of $r$ or $\tau$. The dashed lines are the asymptotic approximations for $g_W \to 1$ calculated through Eq. (\ref{eq:Lambda2_large_g}), relevant to vanishingly small $r$ and $\tau$.}
\label{fig:clipping}
\end{center}
\end{figure}

In principle the function $\Lambda_2[\alpha,\beta,g]$ can be calculated as two-dimensional integral of a bivariate Gaussian distribution. Based on Dirichlet's representation of Heaviside's step function, it can be calculated in the following simpler way\cite{Berk:1991,Teubner:1991,Roberts:1995,Levitz:1998}
\begin{eqnarray} \label{eq:Lambda2}
\Lambda_2[\alpha, \beta, g] &=& \Lambda_1[\alpha] \Lambda_1[\beta] 
+ \frac{1}{2 \pi} \int_0^{\textrm{asin}[g]}  \exp\left[ - \frac{\alpha^2 + \beta ^2 - 2 \alpha \beta \sin(\theta)}{2 \cos^2(\theta)} \right] \ \textrm{d}\theta
\end{eqnarray}
which requires numerically evaluating only a one-dimensional integral. For any given volume fraction of the phases, that is for given $\alpha$ and $\beta$, Eqs. (\ref{eq:Coo}), (\ref{eq:Css}) and (\ref{eq:Caa}) define non-linear relations between the field correlation $g_W$ and the corresponding covariances. These relations are illustrated in Fig. \ref{fig:clipping} for {the centred covariances} $\bar C_{oo}$, $\bar C_{aa}$ and $\bar C_{ss}$, for the values of $\alpha$ and $\beta$ relevant to the microemulsion data. Note that the values satisfy $\alpha=-\beta$, corresponding to $\phi_o=\phi_a$ and $\bar C_{oo}(r)= \bar C_{aa}(r)$.

As visible in Fig. \ref{fig:clipping}, for small field correlations $g_W$ the centred covariances $\bar C_{oo}$ and $\bar C_{aa}$ are proportional to $g_W$. In that region - corresponding to large $r$ or $\tau$ - the bivariate error function $\Lambda_2$ is approximated by 
\begin{eqnarray} \label{eq:Lambda2_small_g}
\Lambda_2[\alpha,\beta,g] &\simeq & \Lambda_1[\alpha] \Lambda_1[\beta] 
+ \frac{g}{2 \pi} \exp\left(-\frac{\alpha^2 + \beta^2}{2} \right)  + \ldots
\end{eqnarray}
which is the first term of a general development in terms of Hermite polynomials\cite{Lantuejoul:2002,Gommes:2013}. However, in general $\Lambda_2$ is a non-linear function. In particular the relation between $g_W$ and the covariances of any clipped structure is vertical when $g_W$ approaches 1 (see Fig. \ref{fig:clipping}). The following asymptotic relation is useful for further purposes
\begin{eqnarray}  \label{eq:Lambda2_large_g}
\Lambda_2[\alpha,\beta,1 - \epsilon^2 ] = \Lambda_1\left[ \textrm{max}\{\alpha,\beta \} \right] &-& 
\frac{\epsilon}{\pi \sqrt{2}} e^{-\frac{\alpha \beta}{2}} e^{-\frac{(\alpha - \beta)^2}{4 \epsilon^2}}  \cr
+ \frac{|\alpha - \beta|}{2 \sqrt{2 \pi}} e^{-\frac{\alpha \beta}{2}} \Big( 1 &-& \textrm{erf} \left[ \frac{|\alpha - \beta|}{2 \epsilon} \right] \Big)
\end{eqnarray}
It is obtained by setting $g=1-\epsilon^2$ in Eq. (\ref{eq:Lambda2}), through a first-order expansion in $\epsilon$. This equation controls the shape of the covariances for asymptotically small $r$ and $\tau$, and therefore the asymptotic shape of the scattering functions for large $q$ and small $\tau$, as we discuss in detail later. The centred covariances approximated through Eq. (\ref{eq:Lambda2_large_g}), are shown as dashed lines in Fig. \ref{fig:clipping}.

\section{Time-dependent clipped Gaussian-field models}
\label{sec:timedependentmodels}

We now introduce three qualitatively different dynamic models to construct time-dependent Gaussian fields, starting from any static Gaussian field. This is achieved by adapting Eqs. (\ref{eq:W_def}) or (\ref{eq:W_def_dilution}), by which the fields are constructed. Although the models are quite general, the discussion is centred on static field GRF-2 of Tab. \ref{tab:fields}, which has the following spectral density
\begin{equation} \label{eq:squaredexponential_f}
f_W(q) =  \left( \frac{l}{2 \sqrt{\pi}} \right)^3   \exp \left[-\frac{(q l)^2}{4} \right] 
\end{equation}
and field correlation function
\begin{equation} \label{eq:squaredexponential_g}
g_W(r) = \exp\left[ -\left(\frac{r}{l} \right)^2 \right]
\end{equation}
where $l$ is model parameter that coincides with the characteristic length $l_W$. With this specific field the main results can be expressed in analytical form. Moreover the corresponding elementary wave $w(\mathbf{x})$ is also known analytically (see Tab. \ref{tab:fields}), which enables one to construct realizations and visually illustrate all considered dynamic models.

All results of Sec. \ref{sec:static} remain valid for time-dependent Gaussian fields. This is notably the case for the clipping relations between the field correlation and the covariances of the water, oil and surfactant phases. However, the field correlation function describes here the statistical correlation between the values of $W$ at two points at distance $r$ apart, with a time lag $\tau$, namely 
\begin{equation} \label{eq:gW_rtau}
g_W(r,\tau) = \langle W(\mathbf{x},t) W(\mathbf{x}+\mathbf{r}, t + \tau) \rangle
\end{equation}
In this case, the covariances obtained through Eqs. (\ref{eq:Coo}-\ref{eq:Caa}) are Van-Hove correlation functions,\cite{Sivia:2011,Squires:2012} and the intensity obtained subsequently through Eqs. (\ref{eq:C_rho} - \ref{eq:I_Fourier}) is the coherent intermediate scattering function $I(q,\tau)$.

Throughout this section, the qualitative geometrical properties of the Gaussian fields are illustrated by clipping them at the value $\alpha=0$. This yields two-phase morphologies with volume fractions $\phi=1/2$, different from the emulsion in Fig. \ref{fig:data}. All the  mathematical results, however, are quite general and remain valid for any clipping procedure. The specific case of the three-phase emulsion, with finite surfactant volume, is considered again in the discussion section.

\subsection{Dynamic model 1: independent time and space fluctuations}
\label{sec:model1}

In the first approach, a field is created with statistically-independent space and time fluctuations. This is achieved by starting from a series of independent static fields $W_n(\mathbf{x})$, with $n=1, \ldots, N$, and combining them linearly with time-dependent coefficients. The fields $W_n$ can be thought of as independent realisations of Eq. (\ref{eq:W_def}) each with different random numbers but the same spectral density $f_W(q)$. The statistical independence is expressed as
\begin{equation} \label{eq:W_def_model1}
\langle W_n(\mathbf{x}) W_m(\mathbf{x}+\mathbf{r}) \rangle =  g_W(r) \delta_{mn}
\end{equation}
where $\delta_{mn} = 1$ for $m=n$ and 0 otherwise.  Based on the set of $W_n(\mathbf{x})$ the time-dependent field is built as
\begin{equation} \label{eq:W_model1}
W(\mathbf{x},t) = \sqrt{\frac{2}{N}} \sum_{n=1}^N W_n(\mathbf{x}) \cos(\omega_n t - \varphi_n)
\end{equation}
where the phases $\varphi_n$ are random and uniform over $[0,2\pi)$, and the frequencies $\omega_n$ are drawn from a temporal spectral density $f'(\omega) \textrm{d}\omega$. With this first dynamic model, the space and time field correlation function in Eq. (\ref{eq:gW_rtau}) is found to be 
\begin{equation} \label{eq:gW_model1}
g_W(r,\tau) = g_W(r) g'(\tau) 
\end{equation}
in the limit of large $N$, with
\begin{equation} \label{eq:gW_time}
g'(\tau) =  \int_0^\infty \cos[\omega \tau] f'(\omega) \textrm{d}\omega 
\end{equation}
In geostatistics, models satisfying Eq. (\ref{eq:gW_model1}) are referred to as being separable.\cite{Gelfand:2010}

A physical interpretation of separable models is obtained by noting that they can be constructed in mathematically-equivalent way through a dilution approach, as in Eq. (\ref{eq:W_def_dilution}). Indeed, the field constructed as
\begin{equation}
W(\mathbf{x},t) = \sqrt{2} \sum_s A_s w(\mathbf{x}-\mathbf{x}_s) \cos(\omega_s t -\varphi_s)
\end{equation}
with $\varphi_s$ uniformly distributed in $[0, 2\pi)$, and $\omega_s$ distributed according to temporal spectral density $f'(\omega)$, has the same correlation function as in  Eq. (\ref{eq:gW_model1}). Therefore, dynamic model 1 can be interpreted as resulting from incoherently fluctuating elementary waves.

\begin{figure}
\begin{center}
\includegraphics[width=6cm]{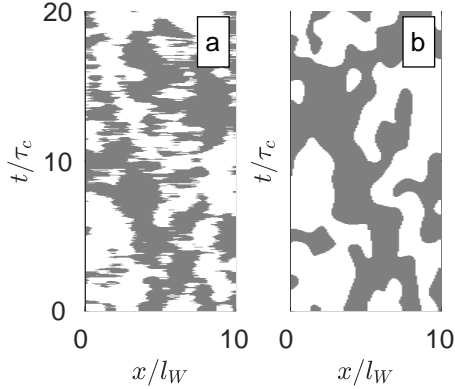}
\caption{Two-dimensional $(x,t)$ cuts through four-dimensional $(x,y,z,t)$ realizations of dynamic model 1, with static field GRF-2 from Tab. \ref{tab:fields} and exponential (a) and hyperbolic secant (b) temporal correlation functions. The threshold assumed in the figure is $\alpha = 0$, corresponding to $\phi=0.5$ for the phases shown in white and grey.}
\label{fig:model1_realizations}
\end{center}
\end{figure}

\begin{figure}
\begin{center}
\includegraphics[width=10cm]{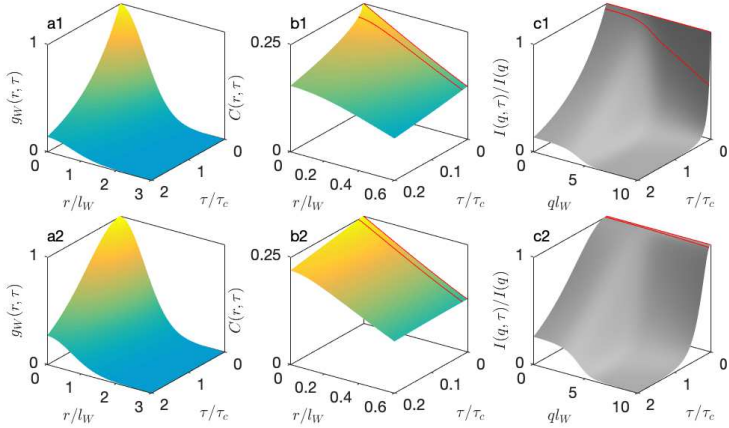}
\caption{Correlation function $g_W(r,\tau)$ for static field GRF-2 and independent space and time fluctuations, with exponential (a1) and hyperbolic-secant (a2) time-correlation functions. The corresponding covariance (clipping threshold $\alpha=0$) and intermediate scattering function are shown in b1/b2 and c1/c2. The two solid lines in b1/b2 and c1/c2 highlight the values for $\tau = 0$ and for small yet finite $\tau$.}
\label{fig:model1}
\end{center}
\end{figure}

For the purpose of data modelling, a natural choice for the temporal correlation function is the exponential $g'(\tau) = \exp[-\tau/\tau_c]$, where the correlation time $\tau_c$ is a model parameter. This choice corresponds to the following spectral density
\begin{equation}
f'(\omega) = \frac{1}{\pi} \frac{2 \tau_c}{1 + (\omega \tau_c)^2}
\end{equation}
A realization of the clipped Gaussian field obtained from this specific time-correlation function and static field GRF-2 is shown in Fig. \ref{fig:model1_realizations}a. The corresponding correlation function $g_W(r,\tau)$ is plotted in Fig. \ref{fig:model1}, together with the covariance assuming a single clipping threshold $\alpha=0$, and the corresponding intermediate scattering function in the form of $I(q,\tau)/I(q)$. 

The type of dynamics obtained from the exponential time correlation function in Fig. \ref{fig:model1_realizations}a is extremely rugged. Smoother dynamics is obtained by modelling the temporal correlation function as a hyperbolic secant $g'(\tau) = 1/\cosh[\tau/\tau_c]$, which behaves like an exponential for asymptotically large times but differs for short times. Its temporal spectral density is
\begin{equation}
f'(\omega) = \frac{2 \tau_c \cosh[\pi \omega \tau_c/2]}{1 + \cosh[\pi \omega \tau_c]}
\end{equation}
A realization of the clipped Gaussian field obtained with this expression is shown in Fig. \ref{fig:model1_realizations}b. The corresponding correlation function $g_W(r,\tau)$, covariance (for $\alpha=0$), and intermediate scattering function are plotted in Fig. \ref{fig:model1}a2 to \ref{fig:model1}c2. 

\subsection{Dynamic model 2: dispersion relation}
\label{sec:model2}

The second dynamic model introduces correlations between space and time fluctuations, and belongs to the class of non-separable models.\cite{Cressie:1999,Gneiting:2002} This is achieved through a dispersion relation that deterministically assigns a specific temporal frequency $\omega$ to any spatial frequency $q$ of the Gaussian field, and leads to the following generalization of Eq. (\ref{eq:W_def})
\begin{equation} \label{eq:W_model2}
W(\mathbf{x},t) = \sqrt{ \frac{2}{N}} \sum_{n=1}^N \sin\left[ \mathbf{q}_n \cdot \mathbf{x} + \omega(|\mathbf{q}_n|) t - \varphi_n \right]
\end{equation}
where $\omega(|\mathbf{q}|)$ is the dispersion relation, which we assume to be isotropic. Based on the statistical independence of the various components of the field in Eq. (\ref{eq:W_model2}), the field correlation function is calculated as
\begin{equation} \label{eq:gW_model2}
g_W(r,\tau) = \int_0^\infty f_W(q) \cos[\omega(q) \tau] \frac{\sin(qr)}{qr} 4 \pi q^2 \textrm{d}q
\end{equation}
in the limit of asymptotically large $N$.

\begin{figure}
\begin{center}
\includegraphics[width=6cm]{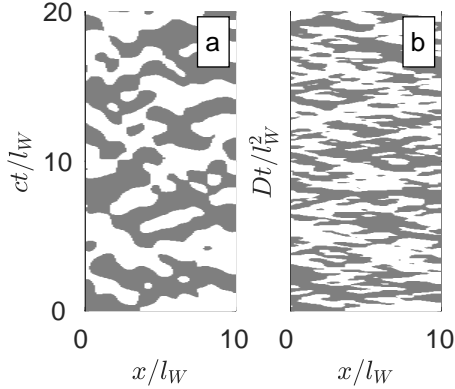}
\caption{Two-dimensional $(x,t)$ cuts through four-dimensional $(x,y,z,t)$ realizations of model 2 of time-dependent Gaussian field, with GRF-2 from Tab. \ref{tab:fields}, and linear (a) and quadratic (b) dispersion relations. The threshold assumed in the figure is $\alpha = 0$, corresponding to $\phi=0.5$ for the phases shown in white and grey.}
\label{fig:model2_realizations}
\end{center}
\end{figure}

In principle any suitable function can be used to model a dispersion relation. We consider here two simple analytical forms $\omega = c q$ and $\omega = D q^2$, where $c$ and $D$ are constants with dimensions of velocity and diffusion coefficient, respectively. Realizations obtained with static field GRF-2 and these two dispersion relations are given in Fig. \ref{fig:model2_realizations}a and \ref{fig:model2_realizations}b. In the case of the linear dispersion relation the structures propagate at constant velocity $c$, which appears as slanted features with slopes $\pm 1$ on the scales of the figure. In the case of quadratic dispersion the structures propagate with size-dependent velocity, which leads to more complicated temporal evolution.

\begin{figure}
\begin{center}
\includegraphics[width=8cm]{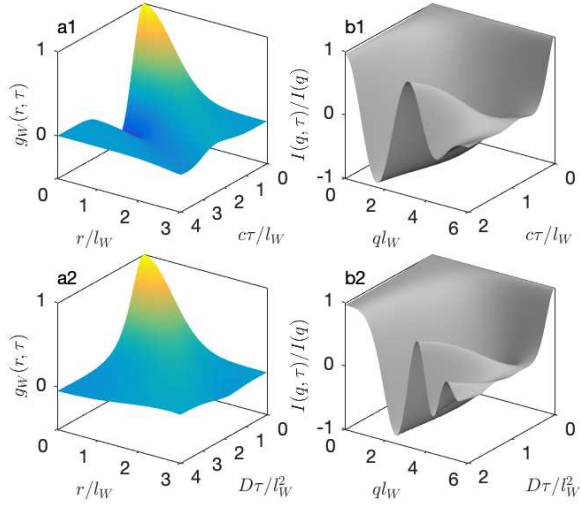}
\caption{Correlation function $g_W(r,\tau)$ for GRF-2 from Tab. \ref{tab:fields} with linear (a1) and quadratic (a2) dispersion relations, together with corresponding intermediate scattering functions (b1 and b2) for clipping threshold $\alpha=0$.}
\label{fig:model2}
\end{center}
\end{figure}

In the particular case of static field GRF-2, analytical expressions are obtained for the field correlation function through Eq. (\ref{eq:gW_model2}). For the linear dispersion relation, one finds
\begin{eqnarray} \label{eq:demo_g_linear}
g_W(r,\tau) = \exp\left[ -\frac{r^2 + (c \tau)^2 }{l_W^2} \right] \times 
\Big\{ 
\cosh\left[ \frac{2 r c \tau }{l_W^2} \right] &-& 2 \left( \frac{c \tau}{l_W} \right)^2 \sinh\left[ \frac{2 r c \tau }{l_W^2} \right]/ \left[ \frac{2 r c \tau }{l_W^2} \right]
\Big\}
\end{eqnarray}
and for the quadratic dispersion, the relation is
\begin{equation} \label{eq:demo_g_quadratic}
g_W(r,\tau) = \frac{\exp\left[ - \frac{(r/l_W)^2}{1 + \left[ 4 D \tau/l_W^2\right]^2} \right]}{\left(1+\left[ 4 D \tau/l_W^2\right]^2 \right)^{3/4}} 
\cos \left[ \frac{(r/l_W)^2 4D \tau/l_W^2}{1 + \left[ 4 D \tau/l_W^2\right]^2} - \frac{3}{2} \tan^{-1}\left[ \frac{4 D \tau}{l_W^2} \right] \right]
\end{equation}
Detailed derivations of these equations are given in the Supplementary Material (Sec. SM-4). The correlation functions and corresponding intermediate scattering functions are plotted in Fig. \ref{fig:model2}. 

An interesting characteristic of the time-dependent structure in Fig. \ref{fig:model2_realizations}a is that it displays temporal order in spite of being spatially disordered. The absence of any feature in $g_W(r,0)$ testifies to spatial disorder (see Fig. \ref{fig:model2}a1 and Fig. SM-3). By contrast, the correlation function $g_W(0,\tau)$ displays a sharp minimum at $\tau = \sqrt{3/2} \times l_W/c$. This corresponds to situation where a fixed point of space is visited alternatively by one phase and the other with quasi periodicity.

The temporal order is not obvious in the correlation function of the quadratic dispersion (Fig. \ref{fig:model2}a2), but the intermediate scattering function $I(q,\tau)$ exhibits marked oscillations for both dispersion relations (Figs. \ref{fig:model2}b1 and b2). This can be understood by noting that for asymptotically large values of $\tau$ the covariance is proportional to $g_W(r,\tau)$ (see the discussion of Fig. \ref{fig:clipping}), so that $I(q,\tau)$ is approximately the Fourier transform of $g_W(r,\tau)$. It therefore results from the Fourier inversion of Eq. (\ref{eq:gW_model2}) that $I(q,\tau)$ is proportional to $f_W(q) \cos[\omega(q) \tau]$ for large values of $\tau$. It is the cosine in this expression that is responsible for the observed oscillations in Fig. \ref{fig:model2}b1 and \ref{fig:model2}b2. For the particular representation as $I(q,\tau)/I(q)$ the oscillations are further amplified by the small value of $I(q)$ for large $q$.

\subsection{Dynamic model 3: moving waves}
\label{sec:model3}

\begin{figure}
\begin{center}
\includegraphics[width=6cm]{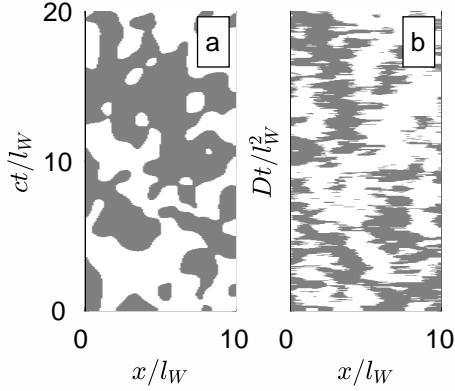}
\caption{Two-dimensional $(x,t)$ cuts through four-dimensional $(x,y,z,t)$ realizations of dynamic model 3 with static field GRF-2 from Tab. \ref{tab:fields}, and (a) ballistic and (b) diffusive propagation of elementary waves. The threshold assumed in the figure is $\alpha = 0$, corresponding to $\phi=0.5$ for the phases shown in white and grey.}
\label{fig:model3_realizations_cD}
\end{center}
\end{figure}

Experimental scattering functions seldom display the type of marked oscillations obtained with dynamic model 2 and displayed in Fig. \ref{fig:model2}. In dynamic model 3, the temporal correlations and corresponding oscillations are naturally damped through a dilution approach, {\it i.e.} by using Eq. (\ref{eq:W_def_dilution}) instead of Eq. (\ref{eq:W_def}) to describe the Gaussian field. Explicitely, the Gaussian field is made time-dependent by allowing the elementary waves to propagate 
\begin{equation} \label{eq:GRF_model3}
W(\mathbf{x},t) = \sum_s A_s w(\mathbf{x} - \mathbf{x}_s - \mathbf{j}_s(t))
\end{equation}
where $\mathbf{x}_s$ is the initial position of wave $s$, and $\mathbf{j}_s(t)$ is the vectorial distance it has travelled at time $t$. We consider two qualitatively different cases: the ballistic or diffusive motions of waves with velocity $c$ or diffusion coefficient $D$. For static field GRF-2 the shape of the elementary wave $w(\mathbf{x})$ is known mathematically (see Tab. \ref{tab:fields}), which enables one to construct realizations as shown in Fig. \ref{fig:model3_realizations_cD}a and b. The diffusive model displays the same type of rugged dynamics as in Fig. \ref{fig:model1_realizations}a, which we analyze in detail in the discussion section.

The field correlation function corresponding to Eq. (\ref{eq:GRF_model3}) is calculated from the time-dependent distributions $f_t(\mathbf{j}) \textrm{d}V_j$ of the wave position $\mathbf{j}$ as follows 
\begin{equation} \label{eq:integral_j}
g_W(\mathbf{r}, \tau )= \theta \langle A^2 \rangle \int   K(\mathbf{r} - \mathbf{j}) f_{\tau}(\mathbf{j}) \ \textrm{d}V_j
\end{equation}
which generalizes Eq. (\ref{eq:gW_dilution}) to time-dependent dilution processes.  If the elementary waves are compact in real space, then $K(\mathbf{r})$ has a range comparable to the characteristic length $l_W$ of the field. It therefore results from Eq. (\ref{eq:integral_j}) that all correlations disappear as soon as the elementary waves have travelled a distance comparable to $l_W$, which happens in a time $l_W/c$ or $l_W^2/D$ for the ballistic or diffusive case, respectively. 

\begin{figure}
\begin{center}
\includegraphics[width=8cm]{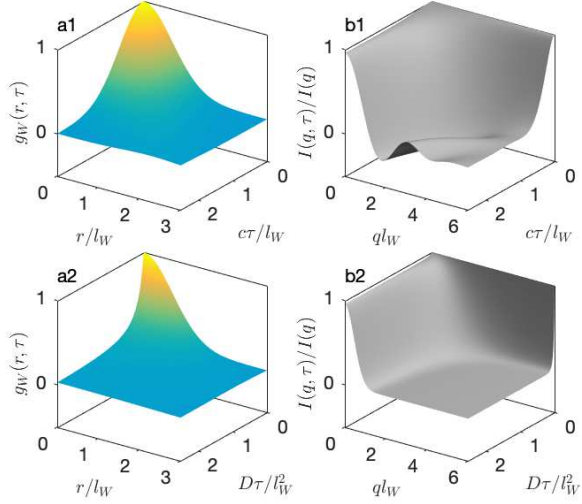}
\caption{Correlation function $g_W(r,\tau)$ of model 3 for Gaussian fields GRF-2 with (a1) ballistic and (a2) diffusive motion of elementary waves, together with corresponding intermediate scattering functions (b1 and b2) for clipping threshold $\alpha=0$.}
\label{fig:model3_cD}
\end{center}
\end{figure}

The case of ballistic propagation corresponds to density distribution
\begin{equation} \label{eq:fj_ballistic}
f_\tau(\mathbf{j}) = \frac{\delta(j - c \tau)}{4 \pi j^2}
\end{equation}
where $j =|\mathbf{j}|$, $\delta(.)$ is Dirac's function, and the denominator accounts for the normalization of the probabilities. With such distribution, Eq. (\ref{eq:integral_j}) reduces to an integration on the unit sphere, and leads to
\begin{equation} \label{eq:gW_model3_c}
g_W(r,\tau)= \theta \langle A^2 \rangle \frac{1}{2} \int_{-1}^{+1} K\left( \sqrt{r^2 + (c \tau)^2 - 2rc\tau \mu}\right) \ \textrm{d}\mu
\end{equation}
In the case of static field GRF-2, this can be calculated explicitly as
\begin{equation}
g_W(r, \tau)=  \exp\left[ - \left(\frac{r - c\tau}{l_W} \right)^2 \right] \frac{1 - \exp\left[ -4 r c \tau /l_W^2 \right]}{4 r c \tau /l_W^2}
\end{equation}
which is plotted in Fig. \ref{fig:model3_cD}a1. The corresponding intermediate scattering function (assuming $\alpha=0$) is plotted in Fig. \ref{fig:model3_cD}b1. 

In the diffusive case, the probability distribution of $\mathbf{j}$ is given by the classical expression for the position of a random walker\cite{Berg:1993,Cussler:2009}  
\begin{equation} \label{eq:fj_diffusive}
f_t (\mathbf{j}) = \left( 4 \pi Dt \right)^{-3/2} \exp\left[ -\frac{|\mathbf{j}|^2}{4Dt} \right]
\end{equation}
For the case of static field GRF-2, Eq. (\ref{eq:integral_j}) then leads to the following field correlation function
\begin{equation} \label{eq:model3_D}
g_W(r,\tau) = \left( 1 + \frac{4 D \tau}{l_W^2} \right)^{-3/2} \exp\left[ -\frac{r^2}{l_W^2+4D\tau} \right]
\end{equation}
This function is plotted in Fig. \ref{fig:model3_cD}a2, with the corresponding intermediate scattering function (assuming $\alpha=0$) in Fig. \ref{fig:model3_cD}b2. 

The asymptotic behaviour of the intermediate scattering function $I(q,\tau)$ for large $\tau$ can be understood by expressing the field correlation function in Eq. (\ref{eq:integral_j}) in terms of the spectral density as follows 
\begin{equation} \label{eq:gW_model3_Fj}
g_W(r,\tau) = \int_0^\infty f_W(q) F_\tau(q) \frac{\sin[q r]}{qr}  4\pi q^2 \textrm{d}q
\end{equation}
where $F_\tau(q)$ is the Fourier transform of $f_\tau(\mathbf{j})$. This expression results directly from Eq. (\ref{eq:integral_j}) by equating $\theta \langle A^2 \rangle K(r)$ to the Fourier transform of $f_W(q)$. For the same reason as for model 2, the intermediate scattering function is  proportional to the Fourier transform of $g_W(r,\tau)$ in the limit of asymptotically large $\tau$. In the case of model 3, this implies $I(q,\tau) \simeq f_W(q) F_\tau(q)$. In the case of the ballistic motion described by Eq. (\ref{eq:fj_ballistic}), the relevant value of $F_\tau(q)$ is
\begin{equation} \label{eq:Fj_c}
F_\tau(q) = \frac{\sin[qc \tau]}{q c \tau}
\end{equation}
which explains the mild oscillations in Fig. \ref{fig:model3_cD}b1. In the case of the diffusive motion described by Eq. (\ref{eq:fj_diffusive}), the relevant function is
\begin{equation} \label{eq:Fj_D}
F_\tau(q) = \exp\left[ - q^2 D \tau \right]
\end{equation}
which explains why no oscillations are observed at all in Fig. \ref{fig:model3_cD}b2. Finally, note that the expression of the field correlation function in Eq. (\ref{eq:gW_model3_Fj}) relies on the spectral density and makes no explicit reference to the elementary wave used to build the model. Therefore the usability of dynamic model 3 is not limited to static Gaussian fields for which the form of the elementary wave is known explicitly.

\section{Discussion}

\subsection{Temporal crossing rate}
\label{sec:crossing_rates}

An interesting characteristic of the Neutron Spin-Echo (NSE) data in Fig. \ref{fig:data}c1 and \ref{fig:data}c2 is the very steep $\tau$-dependence for small $\tau$ and large $q$. Among the three dynamic models discussed in Sec. \ref{sec:timedependentmodels}, this type of behavior was observed for model 1 with exponential time correlation function (Fig. \ref{fig:model1}c1), and for model 3 with diffusive motion of elementary waves (Fig. \ref{fig:model3_cD}b2). In both cases, the realisations testify to extremely rugged dynamics as illustrated in Figs. \ref{fig:model1_realizations}a and \ref{fig:model3_realizations_cD}b. 

A useful mathematical concept to describe the two types of dynamics in both Fig. \ref{fig:model1_realizations} and Fig. \ref{fig:model3_realizations_cD} is the temporal crossing rate $n_t$, which characterizes how often a fixed point in space is crossed by moving interfaces. As discussed shortly, this concept is the temporal equivalent of the spatial notion of specific surface area. The surface area $a_V$ is defined as the total area of an interface per unit volume of the system. For an isotropic system it is mathematically related to the notion of average chord length, which characterizes how frequently one crosses the interface when traveling along any straight line crossing the system.\cite{Gommes:2020B} The significance of the surface area for scattering was first acknowledged by Debye, who related $a_V$ to the small-$r$ behaviour of the covariance as\cite{Debye:1957} 
\begin{equation} \label{eq:Debye}
C(r,0) \simeq \phi - \frac{a_V}{4} r + \ldots
\end{equation}
which converts in reciprocal space to the well-known Porod's law\cite{Guinier:1963,Ciccariello:1988,Ciccariello:1995}
\begin{equation} \label{eq:Porod}
I(q,0) \simeq \frac{2\pi a_V}{q^4}
\end{equation}
In the particular case of clipped Gaussian-field structures, the surface area of an isosurface, say at $W(\mathbf{x})=\alpha$, is calculated as\cite{Teubner:1991,Berk:1991}
\begin{equation} \label{eq:surface_area}
a_V= \frac{2\sqrt{2}}{\pi l_W} e^{-\alpha^2/2}
\end{equation}
where $l_W$ is the characteristic length of the field defined in Eq. (\ref{eq:lW}). 

\begin{figure}
\begin{center}
\includegraphics[width=7cm]{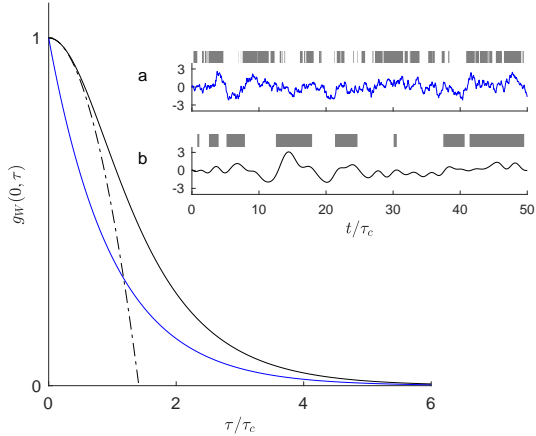}
\caption{Field correlation functions at a fixed point in space $g_W(0,\tau)$ for dynamic model 1 with exponential (blue) and hyperbolic-secant (black) temporal correlation functions. The insets are realizations of the time-dependent Gaussian fields at a fixed point in space, together with clipped structure with $\alpha=0$, for the exponential (a) and hyperbolic-secant (b) models. The dashed line highlights the quadratic shape of the hyperbolic-secant model at the origin.}
\label{fig:excursion_1D}
\end{center}
\end{figure}

The crossing rate $n_t$ is related to the covariance via
\begin{equation} \label{eq:C_nt}
C(0,\tau) = \phi - \frac{n_t}{2} \tau + \ldots
\end{equation}
which illustrates further the similarity between $n_t$ and the surface area $a_V$ in Eq. (\ref{eq:Debye}). The factor
 $1/2$ differs from the factor $1/4$ in Eq. (\ref{eq:Debye}) because the temporal process considered here is one-dimensional.\cite{Torquato:2002} From Eq. (\ref{eq:C_nt}) the crossing rate can be calculated as the limit of $-2 (\partial C/\partial \tau)$ for vanishingly small $r$ and $\tau$. In the case of a clipped Gaussian field model, this limit corresponds to values of the field correlation function $g_W(r,\tau)$ asymptotically close to 1. The derivative can then be calculated using the asymptotic result in Eq. (\ref{eq:Lambda2_large_g}). The following expression is then obtained for the crossing rate 
\begin{equation} \label{eq:nt}
n_t = \frac{\sqrt{2}}{\pi} e^{-\alpha^2/2} \lim_{\tau \to 0} \frac{\sqrt{1-g_W(0,\tau)}}{\tau}
\end{equation}
which provides a physical interpretation to the small-$\tau$ behavior of $g_W(0,\tau)$. The condition for having finite $n_t$ is that the field correlation function should be quadratic at the origin
\begin{equation}
g_W(0,\tau) \simeq 1 - (\tau/\tau_W)^2 + \ldots
\end{equation}
which defines a natural characteristic time $\tau_W$. Note that the two dynamic models of Sec. \ref{sec:timedependentmodels} displaying rugged dynamics both have correlation functions that are linear at the origin. In the case of model 1 with exponential correlation function, Eq. (\ref{eq:gW_model1}) leads to
\begin{equation}
g_W(0,\tau) = 1 - \tau/\tau_c + \ldots
\end{equation}
and in the case of model 3 with diffusive wave motion, Eq. (\ref{eq:model3_D}) leads to
\begin{equation} \label{eq:gW_tau_D}
g_W(0,\tau) = 1 - \frac{6 D \tau}{l_W^2} + \ldots
\end{equation}
In both cases $\tau_W$ is not defined and Eq. (\ref{eq:nt}) predicts infinite crossing rate. The effect of linear versus quadratic correlation at short times is illustrated further in Fig. \ref{fig:excursion_1D}.

Whether the crossing rate $n_t$ is finite or infinite controls the shape of the intermediate scattering function $I(q,\tau)$ for asymptotically large $q$ and small $\tau$ (see Figs. \ref{fig:model1} and \ref{fig:model3_cD}). The analysis builds on the following two mathematical facts. (i) First, the non-linearity of the clipping function at $g_W=1$ is of the type $C \simeq 1 - \sqrt{1-g_W}$ (see Fig. \ref{fig:clipping} and Eq. \ref{eq:Lambda2_large_g}). In particular, this converts a quadratic field correlation $g_W(r,\tau) \simeq 1-r^2$ into a linear covariance $C(r,\tau) \simeq 1-r$. This also converts a linear field correlation $g_W(r,\tau) \simeq 1-\tau$ into a singular covariance $C(r,\tau) \simeq 1-\sqrt{\tau}$. (ii) Second, the asymptotic behavior of $I(q,\tau)$ for large $q$ is controlled by the small-$r$ behavior of $C(r,\tau)$. This results from a generalization of the Riemann-Lebesgue lemma by Lighthill\cite{Lighthill:1958,Berk:1991}, and is shortly discussed in the Supplementary Material (Sec. SM-6). In particular, a covariance that is linear at the origin $C(r,0) = 1 - r$ leads to a $1/q^4$ scattering, in line with Porod's law in Eq. (\ref{eq:Porod}). By contrast, a covariance whose derivatives all vanish at $r=0$ leads to a scattering that decreases faster than any power law.  

Consider now the case where $\tau_W$ exists, {\it i.e.} where $g_W(r,\tau)$ is quadratic in $\tau$ at the origin, and $C(r,\tau)$ is linear in $\tau$ ({\it e.g.} Figs. \ref{fig:model1}b2). In that case Porod's law holds not only for $I(q)$ but also for $I(q,\tau)$ for small $\tau$s, because $C(r,\tau)$ is a smooth function of $\tau$. As a consequence $I(q,\tau)/I(q)$ approaches the value 1 horizontally for $\tau \to 0$ (Fig. \ref{fig:model1}c2). By contrast, if $\tau_W$ is not defined the covariance $C(r,\tau)$ varies like $\sqrt{\tau}$, which has infinite slope for $\tau \to 0$. Accordingly $C(r,\tau)$ passes from being linear in $r$ to having vanishing derivative over infinitesimally short interval of $\tau$ (Fig. \ref{fig:model1}b1). The intermediate scattering function $I(q,\tau)$ passes discontinuously from Porod's law for $\tau = 0$ to decreasing faster than $q^{-4}$ for arbitrarily small $\tau > 0$. This explains the very steep intermediate scattering function $I(q,\tau)$ for large $q$ and small $\tau$ in  Fig. \ref{fig:model1}c1. The same explanation holds for Fig. \ref{fig:model3_cD}b2.

In the case of model 1 the existence $\tau_W$ and the finiteness of $n_t$ can be ascertained by direct examination of $g'_W(\tau)$. In the case of model 2, one has to examine both the dispersion relation and the spectral density. Assuming a dispersion relation of the type
\begin{equation}
\omega(q) = a_n q^n
\end{equation}
where $a_n$ and $n$ are constants, a truncated expansion of the cosine factor in Eq. (\ref{eq:gW_model2}) leads to 
\begin{equation} \label{eq:tauW_dispersion}
\frac{1}{\tau_W^2} = \frac{a_n^2}{2} \int_0^\infty f_W(q) q^{2n} 4 \pi q^2 \textrm{d}q
\end{equation}
In the particular case of a linear dispersion relation, with $a_1 = c$, the characteristic time is proportional to the characteristic length $\tau_W = l_W/(c \sqrt{3})$. However, for exponents $n$ larger than one the conditions are more stringent for $\tau_W$ than for $l_W$. The condition for non-vanishing $\tau_W$ is that the spectral density $f_W(q)$ should decrease faster than $q^{-\nu}$ with $\nu = 2 n +3$ (see also Sec. SM-VII in the supporting information) . Finally, in the case of model 3, it results from Eq. (\ref{eq:gW_model3_Fj}) that the ballistic case always leads to finite $n_t$, provided $l_W$ is finite. Replacing Eq. (\ref{eq:Fj_c}) by a truncated expansion for small $\tau$, the following expression is indeed obtained 
\begin{equation}
g_W(0,\tau) \simeq 1 - \left( \frac{c \tau}{l_W} \right)^2 + \ldots
\end{equation}
which shows that $\tau_W = l_W/c$. By contrast, in the diffusive case, $g_W(0,\tau)$ is linear in $\tau$ and $n_t$ is always infinite, as already shown in Eq. (\ref{eq:gW_tau_D}). The latter equation holds for all spectral densities, and it is not limited to static field GRF-2.
 
\subsection{NSE data analysis}
\label{sec:data_analysis}

\begin{figure}
\begin{center}
\includegraphics[width=8cm]{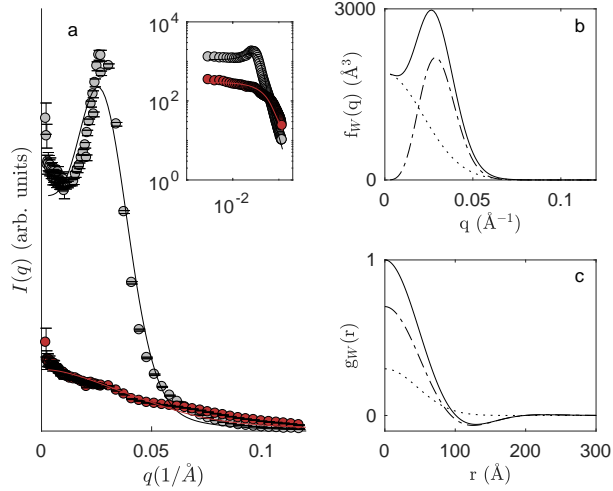}
\caption{(a) Fitting of the microemulsion SANS patterns of Fig. \ref{fig:data} as a sum of two static GRF contributions, with bulk and film contrasts in red and grey, respectively. The dots are the data, and solid lines are the fits. The two contributions to the spectral density and the correlation functions are shown in b and c, with background contribution (GRF-2, dotted), correlated contribution (GRF-4,dashed) and their sum shown as a solid line.}
\label{fig:SANS_twowaves}
\end{center}
\end{figure}

The developed approach offers the possibility of decomposing a given static structure into distinct contributions, and building composite time-dependent models whereby each structural contribution is animated according to different dynamic models. In the particular case of the microemulsion, the static SANS data hint at two types of structures as illustrated in Fig. \ref{fig:SANS_twowaves}. A distinctive feature of the SANS is the presence of a sharp scattering peak in the bulk contrast data around 0.03 \AA$^{-1}$, corresponding to strongly correlated structures. That peak alone, however, does not describe the entire SANS as it is superimposed with a more diffuse and featureless background scattering that extends over the entire $q$ range. We now endeavour to model these two contributions with suitable Gaussian fields, and use these SANS-born models to explore the NSE data.

A natural choice for modelling the background-like contribution in the SANS is the static field GRF-2, which was used for illustrative purposes throughout Sec. \ref{sec:timedependentmodels}. As the spectral density of GRF-2 (see Eq. \ref{eq:squaredexponential_f}) displays no peak, it is not suitable to model the correlated part of the structure. For the latter contribution, we introduce a new static field, the elementary wave of which is built as the Laplacian of GRF-2, namely
\begin{equation} \label{eq:w_Laplacian}
w(\mathbf{x}) = \left[  \left(\frac{ |\mathbf{x}|}{l}\right)^2 - \frac{3}{4} \right] \exp\left[- 2 \left( \frac{|\mathbf{x}|}{l} \right)^2 \right] 
\end{equation} 
The static properties of this field are given in Tab. \ref{tab:fields} under the name GRF-4. By construction the elementary wave in Eq. (\ref{eq:w_Laplacian}) satisfies
\begin{equation} \label{eq:int_0}
\int w(\mathbf{x}) \textrm{d}V_x = 0
\end{equation}
This is equivalent to $f_W(0)=0$ by virtue of Eq. (\ref{eq:fW_w}), which leads to the desired scattering peak in the SAS patterns (See also Fig. SM-4). Interestingly, in the context of moving-wave dynamic models (model 3) when a given volume is crossed by any wave satisfying Eq. (\ref{eq:int_0}) the local average value of the Gaussian field $W(\mathbf{x})$ remains unchanged, in a neighbourhood of size larger than $l_W$. Therefore, the propagation of the elementary wave in Eq. (\ref{eq:w_Laplacian}) preserves locally the volumes of the phases. We therefore refer to GRF-4 as a {\em deformation} mode. The volume-preservation property can also be understood by noting that the low-$q$ limit of the scattered intensity is proportional to the compressibility of the phases \cite{Glatter:1982} so that a spectral density that vanishes for $q \to 0$ corresponds to incompressible phases. By contrast, GRF-2 leads to local modifications of the the volumes, and we refer to it as a {\em breathing} mode. The spectral densities of these two modes are illustrated in Fig. \ref{fig:SANS_twowaves}b.

To analyze the microemulsion SANS data the breathing and deformation modes were combined into a single Gaussian field, following Eq. (\ref{eq:combine_W}). This leads to a three-parameter static-field model, with the characteristic lengths of each mode and their relative contribution to the field variance. The clipping thresholds $\alpha=-0.234$ and $\beta=+0.234$ are imposed by the volume fractions. The least-square fit of both bulk- and film-contrast data is illustrated in Fig. \ref{fig:SANS_twowaves}. The breathing mode contributes 70 \% of the variance ($\sigma_b^2 \simeq 0.7$) with lengths $l_d \simeq $ 98 \AA \ and $l_b \simeq$ 65 \AA \ for the deformation and breathing modes. These numerical values of $l_d$ and $l_b$ coincidentally correspond to the same characteristic lengths $l_W \simeq 65$ \AA (see Tab. \ref{tab:fields}).

\begin{figure}
\begin{center}
\includegraphics[width=14cm]{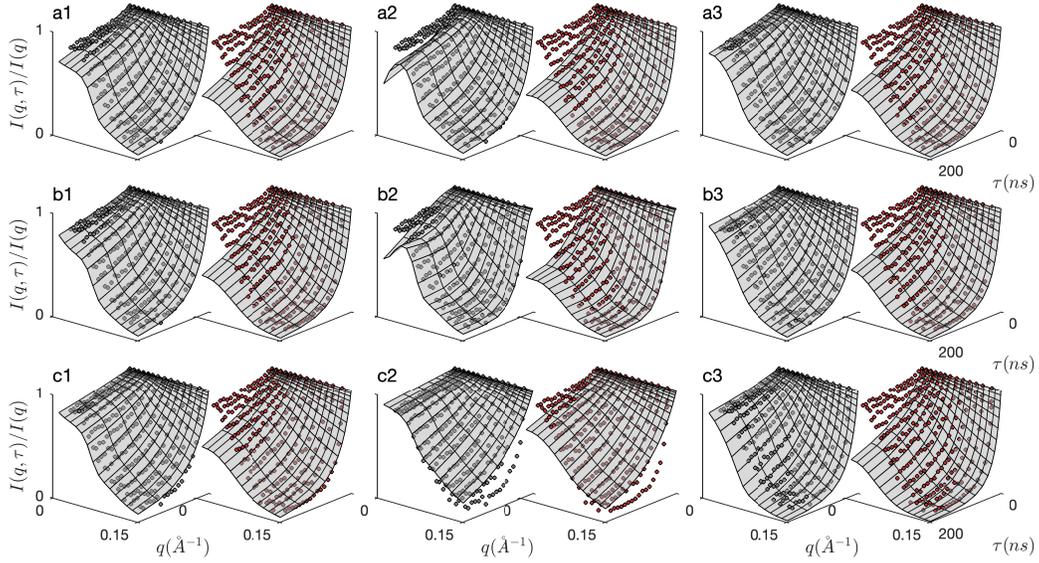}
\caption{Fitting of the microemulsion NSE data with the composite breathing-and-deformation model (GRF-2 and GRF-4), with the dynamics of the modes modelled either as (i) fluctuations,  (ii) ballistic waves, or (iii) diffusive waves. From top to bottom: the breathing mode is assigned fluctuation (a), ballistic (b), or diffusive dynamics (c). From left to right: the deformation mode is assigned fluctuation (1), ballistic (2), or diffusive dynamics (3). In each case the dots are the data with bulk (grey) and film (red) contrasts, and the surface is the model. The corresponding parameters are in Tab. \ref{tab:parameters}.}
\label{fig:NSE_composite}
\end{center}
\end{figure}

In order to analyze the NSE data, a dynamic model has to be assumed for each mode. As the notion of breathing and deformation is inspired by an elementary-wave interpretation we restrict the analysis to model 1 (fluctuating waves) and model 3 (ballistically or diffusively propagating waves). Moreover, as the NSE data exhibit steep slope for large $q$ and small $\tau$ (Fig. \ref{fig:data}c1-c2), we consider only the exponential correlation function for model 1 with infinite crossing rate $n_t$. In the following, we explore systematically all combinations of the three types of dynamics for the two modes, which leads to nine composite time-dependent Gaussian fields. The least-square fits of the NSE data are illustrated in Fig. \ref{fig:NSE_composite}, and the values of the corresponding parameters and $\chi^2$ are reported in Tab. \ref{tab:parameters}. The fitting required the correlation function $g_W(r,\tau)$ to be known for the deformation mode (GRF-4), for both ballistic and diffusive wave propagation. All details are provided in Sec. SM-V of the Supplementary Material.

\begin{table}
\begin{tabular}{c|c|p{4cm}|p{4cm}|p{4cm}| }
\multicolumn{2}{r}{}
 & \multicolumn{3}{c}{Deformation mode} \\
 \cline{3-5}
 \multicolumn{2}{r}{}
 &  \multicolumn{1}{c}{Fluct. ($\tau_d$)}
 & \multicolumn{1}{c}{Ball. ($c_d$)} 
 & \multicolumn{1}{c}{Diff. ($D_d$)} \\
\cline{3-5}
 \parbox[t]{10mm}{\multirow{7}{*}{\rotatebox[origin=c]{90}{Breathing mode}}} 
 & \parbox[t]{8mm}{\rotatebox[origin=c]{90}{Fluct. ($\tau_b$)}} & \makecell{$\tau_b = 1000$ ns$^*$ \\ $\tau_d = 519$ ns \\ ($\chi^2=10$) } & 
\makecell{$\tau_b = 130$ ns \\ $c_d = 0.088$ \AA/ns \\ ($\chi^2=11$) } & 
\makecell{$\tau_b = 1000$ ns$^*$ \\ $D_d = 1.3$ \AA$^2$/ns \\ ($\chi^2=8.2$) } \\
\cline{3-5}
 & \parbox[t]{8mm}{\rotatebox[origin=c]{90}{Ball. ($c_b$)} }& \makecell{$c_b = 0.001$ \AA/ns$^*$ \\ $\tau_d = 429$ ns \\ ($\chi^2=9.6$) } &
\makecell{$c_b = 0.82$ \AA/ns \\ $c_d = 0.003$ \AA/ns \\ ($\chi^2=14$) } & 
\makecell{$c_b = 0.001$ \AA/ns$^*$ \\ $D_d = 1.7$ \AA$^2$/ns \\ ($\chi^2=7.4$) } \\
\cline{3-5}
& \parbox[t]{8mm}{\rotatebox[origin=c]{90}{Diff. ($D_b$)}} & \makecell{$D_b = 4.0$ \AA$^2$/ns \\ $\tau_d = 1000$ ns$^*$ \\ ($\chi^2=8.4$) } &
\makecell{$D_b = 8.6$ \AA$^2$/ns \\ $c_d = 0.081$ \AA/ns \\ ($\chi^2=7.6$) } & 
\makecell{$D_b = 0.001$ \AA$^2$/ns$^*$ \\ $D_d = 1.7$ \AA$^2$/ns \\ ($\chi^2=7.4$) } \\
\cline{3-5}
\end{tabular}
\caption{Parameters of the composite model, whereby each mode (breathing or deformation, both with $l_W \simeq 65$ \AA) is assigned either a fluctuating, ballistic or diffusive dynamics. The values were obtained from the least-square fits in Fig. \ref{fig:NSE_composite}, and the $\chi^2$ values are also reported. The stars highlight values that have converged to the lower bound allowed for the fit.}
\label{tab:parameters}
\end{table}

Globally, none of the composite models in Fig. \ref{fig:NSE_composite} is able to quantitatively account for both the bulk- and film-contrast NSE data. The model that fares worst is the one that assumes ballistic wave propagation for both breathing and deformation modes. This model has finite crossing rate $n_t$, and is unable to capture the steep experimental intermediate scattering function. Based on the $\chi^2$ values in Tab. \ref{tab:parameters}, the data is best described when both modes have diffusive dynamics. More generally, it is interesting to note that the calculated intermediate scattering functions in Fig. \ref{fig:NSE_composite} are generally smaller than the data, particularly for the film-contrast scattering, so that the breathing and deformation-mode approach overestimates the dynamics. This is also manifest in the values of Tab. \ref{tab:parameters}, which often converge towards the lower limits allowed on the parameters. This means that one of the two modes is practically static and does not contribute to the dynamics.

As an alternative to the composite model based on breathing and deformation modes, the stochastic models offer the possibility of a more general approach based on a dispersion relation (dynamic model 2). This enables one to better match experimental data by tuning the dynamics in a scale-dependent way through the value of $q$. Moreover, the dispersion-relation approach does not require one to explicitly decompose the SANS into substructures, so that the same piecewise-linear spectral density (GRF-8) as in Fig. \ref{fig:data}a and SM-10 can be used to describe the underlying static Gaussian field. 

\begin{figure}
\begin{center}
\includegraphics[width=8cm]{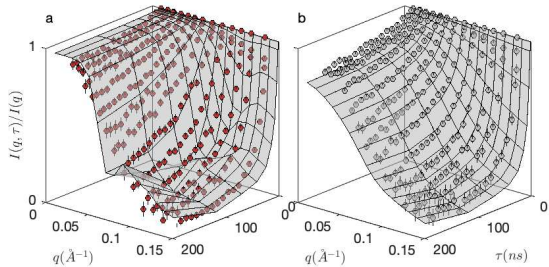}
\caption{Least-square fit of the NSE data from Fig. \ref{fig:data} with a dispersion relation (dynamic model 2) and a static field with piecewise-linear spectral density (GRF-8, same as Fig. \ref{fig:data}a), for both bulk contrast (a) and film contrast (b). The dispersion relation is Eq. (\ref{eq:dispersion}) with parameter $a \simeq 2053\ \AA^3$/ns and $q_c \simeq 0.047\  \AA^{-1}$, resulting in $\chi^2 = 6.3$. The surfaces are the model and the dots are the data, with errorbar $\pm 2 \sigma$.}
\label{fig:NSE_dispersion}
\end{center}
\end{figure}

Although any dispersion relation can in principle be chosen to model the NSE data, it is desirable that it should lead to an infinite crossing rate $n_t$ so as to capture the steep intermediate scattering function. As discussed in Sec. SM-III of the Supplementary Material, the spectral density of the piecewise-linear model decreases asymptotically as $q^{-6}$ so that any dispersion relation with order $n>3/2$ would lead to infinite $n_t$. In practice, the following dispersion relation is found to describe fairly the NSE data
\begin{equation} \label{eq:dispersion}
\omega = a (q-q_c)^3 H[q-q_c]
\end{equation}
where $H()$ is Heaviside's step function, $q_c$ is a cutoff frequency, and the parameter $a$ sets the value of $\omega$. The least-square fit is illustrated in Fig.\ref{fig:NSE_dispersion} and in Fig. SM-12 as 2D plots. The values of the fitted parameters are $a \simeq 2050 \pm 60 \ \AA^3$/ns and $q_c \simeq 0.047 \pm 0.0002 \  \AA^{-1}$, resulting in $\chi^2 = 6.3$, which is a significant improvement compared to the values in Tab. \ref{tab:parameters}. The reported uncertainties on $a$ and $q_c$ were obtained from a Monte-Carlo estimation, with a $\pm 2 \sigma $ normal-distributed error on each NSE data point. The small errors on the parameters results from the fact that all the NSE data - over the entire $q$ and $\tau$ ranges, and for the two contrasts - are fitted jointly with only two adjustable parameters which makes it quite robust. This also offers the prospect of reducing the number of experimental data points needed to reliably adjust a model (see Fig. SM-15).

As an alternative to Eq. (\ref{eq:dispersion}) the NSE data were also fitted with a dispersion relation modelled as a sum of power laws (from $n=1$ to $n=4$) with adjustable factors. Such fit converges to a situation where factors with alternating signs contribute to keeping $\omega$ close to zero for small $q$, leading to an overall shape similar to the cutoff frequency used in Eq. (\ref{eq:dispersion}) (see Fig. SM-13). 

The $q^3$ dependence assumed in Eq. (\ref{eq:dispersion}) appears in a variety of dynamic structure factors involving hydrodynamic interactions, although the specifics of the relaxation curve are system-dependent. The exponent $3$ notably appears in the Zimm model of polymers in solvents to describe the thermally driven fluctuations of a polymer chain \cite{Zimm:1956,Dubois:1967}. It also appears in the Zilman-Granek analysis of membrane fluctuations with bending rigidity \cite{Zilman:1996}. In the present context this can be understood from the following general scaling argument. The very observation of infinite $n_t$ hints at thermal random motion. By itself, this would lead to a quadratic dispersion relation $\omega = D q^2$, where the diffusion coefficient $D$ is related to the characteristic size $L$ by Stokes-Einstein relation $D \simeq k_B T /(\eta L) $, where $k_BT$ is the thermal energy and $\eta$ is the viscosity of the medium. In the case of microemulsion deforming over a variety of length scales, inversely related to the scattering vector $L \sim q^{-1}$, this suggests the following cubic dispersion relation
\begin{equation}
\omega \sim \frac{k_B T}{\eta} q^3
\end{equation}
Using the value $\eta \simeq 10^{-3}$ Pa.s for the effective viscosity of the water/decane microemulsion as reported in ref.\cite{Holderer:2005} the factor in this scaling law takes the value 4000 \AA$^{3}$/ns, which is of the same order of magnitude as the value inferred from the NSE data fitting.

\begin{figure}
\begin{center}
\includegraphics[width=12cm]{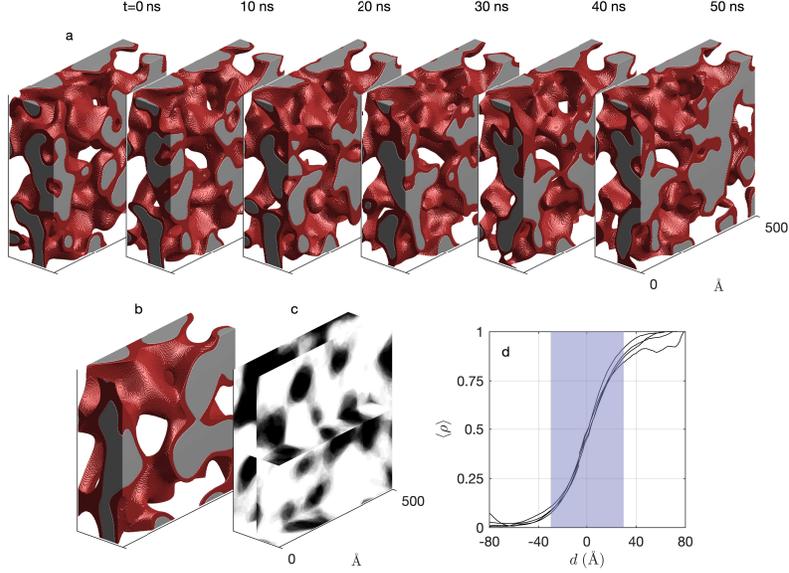}
\caption{(a) Time-dependent realizations of the model of Fig. \ref{fig:NSE_dispersion} over 50 ns, with oil in grey, surfactant in red, and water in white; (b) realization from the static components of the field only, corresponding to spectral density limited to $q < q_c$; (c) average oil density calculated over the duration of the simulation; (d) average oil density profile as a function of distance to the interface of the static components ($d$ is positive into the oil). The blue area highlight the 80 \% confidence interval, with width $\simeq 60$ \AA. The four curves in d were obtained from independent realizations.}
\label{fig:realization_versus_time}
\end{center}
\end{figure}

The time-dependent structure of the microemulsion is illustrated in Fig. \ref{fig:realization_versus_time} with a particular realization of the dispersion-relation model over a time interval of 50 ns. One notes in particular the stability in time of large-scale structures, larger than approximately $2 \pi/q_c \simeq 130$ \AA, where $q_c$ is the cutoff frequency from Eq. (\ref{eq:dispersion}). The average position of the oil and water phases does not change over the timescale of the figure. However, the interfaces deform in random and very rugged way at smaller scale, as typically expected from a system with infinite crossing rate $n_t$. This picture matches the physical intuition as the redistribution of oil and water over long distances occurs through slow hydrodynamic flow, while no such obstacle exists for the local fluctuations. This scale-dependence is captured by the Stokes-Einstein relation, and is responsible for the observed $q^3$ dynamics. On the other hand, the physical origin of the cutoff size $q_c$ remains unclear, although its phenomenology is reminiscent of de-Gennes narrowing, whereby relaxation times are often observed to scale with the scattering structure factors\cite{Myint:2021}.

In order to better understand the realizations of the fitted model, the structure was further decomposed into its static and time-dependent components. The realisation in Fig. \ref{fig:realization_versus_time}b was obtained by setting to zero all components of the spectral density $f_W(q)$ with $q> q_c$, which results in much smoother interface. This is manifest in the characteristic lengths of the field (Eq. \ref{eq:lW}), which pass from $l_W \simeq 52$ \AA \ to $l_W \simeq 73$ \AA. Based on Eq. (\ref{eq:surface_area}) this corresponds to surface areas $a_V \simeq 170$ m$^2$/cm$^3$ and $a_V \simeq 120$ m$^2$/cm$^3$, respectively.  The thermal fluctuations of the interface therefore contribute to as much as 30 \% of the area of the interfaces. Due to the symmetry of the model (with clipping constants $\alpha=-\beta$) the  surfactant/oil and surfactant/water interfaces have identical areas. Another interesting aspect of the fluctuations is their amplitude, which can be estimated by evaluating first the average density, say, of oil over the entire duration of a simulation. This is illustrated in Fig. \ref{fig:realization_versus_time}c, where the smooth transition between the white and black areas correspond to all the successive positions of the interface over time. The extent of the transition in the direction locally orthogonal to the interface is given in Fig. \ref{fig:realization_versus_time}d. During 80 \% of the time, the interface fluctuates within a 60 \AA-thick layer that extends on both sides of the average position. It is interesting to compare that value to the size of the oil and water phases, estimated as an average chord length\cite{Gommes:2020B} as $4 \phi/a_V \simeq 130 \ \textrm{\AA}$ for the average structure in Fig. \ref{fig:realization_versus_time}b. In other words, the interface fluctuates over distances as large as half the size of the phases.

The dispersion-relation analysis hints at reasons why the breathing and deformation-mode analysis was unable to account for the NSE data of the emulsion. The dispersion relation points indeed at two dynamic regimes but they are separated by a cutoff frequency, which is much sharper a transition than between GRF-2 and GRF-4. It has also to be noted that the very concept of independent modes contributing additively to the dynamics, is strictly justified only as a linear approximation. Given the observed large amplitude of the interface fluctuations, non-linear effects could be expected which would rule out any possibility of linear-mode decomposition.

\section{Conclusion}

Clipped Gaussian field models have been extensively used to analyze the elastic small-angle scattering data of disordered systems. When applied to dynamical systems, such classical approach provides static snapshots of a structure. In the paper, the models are generalised to make them time-dependent, which enable one to analyze consistently both the instantaneous spatial structures and their dynamics within a single statistical description. General expressions are derived for all the space- and time-correlation functions relevant to coherent inelastic neutron scattering, for multiphase systems and arbitrary scattering contrasts between the phases.

With the proposed approach, for any given static structure inferred {\it e.g.} from small-angle scattering, a variety of distinctly different dynamics can be modelled. In a first family of models, the Gaussian field underlying the structure is decomposed into a large number of localised elementary waves. Qualitatively different dynamics are obtained by letting the waves randomly fluctuate, or propagate ballistically or diffusively through the system. In another family of models, the spectral components of the Gaussian field are assigned any desired dynamics through a suitable dispersion relation. The various types of dynamics lead to qualitatively different intermediate scattering functions, which enables one to discriminate them through neutron scattering. Moreover, all these approaches can be combined to yield models with composite and possibly realistic dynamics.

A central characteristic of the dynamic models, which controls the shape of the intermediate scattering functions, is their temporal crossing rate. This is defined by considering a fixed point in space, and evaluating how often it is passed through by a moving interface of the time-dependent structure. Systems undergoing Brownian-like thermal fluctuations have infinite crossing rate, which converts to infinitely steep intermediate scattering functions for asymptotically large $q$ and small $\tau$. 

The methods of the paper were illustrated with the analysis of neutron small-angle scattering and spin-echo data measured on oil/water microemulsions. The methodology consisted in analyzing first the SANS data in order to determine the spectral density of the Gaussian field underlying the static structure, corresponding to snapshots of the time-dependent structure. As a second step, the NSE data was analyzed by complementing the so-determined static spectral density with few dynamic parameters. This enabled us to analyze jointly the entire SANS and NSE data, in both film and bulk contrasts and over the entire range of $q$ and $\tau$, with a single coherent model. The small number of adjusted parameter contributes to the robustness of the NSE analysis, and offers the prospect of reducing the number of experimental points required to reliably adjust a model.

From a physical perspective, the SANS and NSE data of the emulsion point to a static large-scale structure of the oil and water domains, with thermal fluctuations of the interfaces. The interface fluctuations take place over distances as large as 60 \AA , corresponding to half the domain size, and contribute to 30\% of the total interface area. In future work the stochastic approach will be explored further to analyze the wavelike dynamics observed by neutron spin-echo in lipid membranes\cite{Jaksch:2017}.  

\section*{Supplementary Material}

See supplementary material for the mathematical derivation of some equations, for numerical data-analysis procedures, as well as for additional figures.

\section*{Data Availability}

All SANS and NSE data discussed in the paper can be downloaded from the authors' institutional repository at \url{https://doi.org/10.26165/JUELICH-DATA/DJ3LIN}.

\nocite{*}
\bibliography{timedependentgrf}

\end{document}